\documentclass[aps,pra,letterpaper,twocolumn,showpacs,floatfix,groupedaddress,superscriptaddress]{revtex4}
\usepackage{graphics,graphicx,bbm,bm,times,color,amsfonts,amssymb,amsmath,natbib}
\bibpunct[]{[}{]}{,}{a}{}{;}

\begin{document}

\title{Direct estimation of single- and two-qubit Hamiltonians and
relaxation rates}
\author{M. Mohseni}
\affiliation{Department of Chemistry and Chemical Biology, Harvard University, 12 Oxford
St., Cambridge, MA 02138, USA}
\author{A. T. Rezakhani}
\affiliation{Center for Quantum Information Science and
Technology, and Departments of Physics and Chemistry, University of
Southern California, Los Angeles, CA 90089, USA}
\affiliation{Institute for Quantum Information Science, University
of Calgary, Alberta T2N 1N4, Canada}
\author{A. Aspuru-Guzik}
\affiliation{Department of Chemistry and Chemical Biology, Harvard
University, 12 Oxford St., Cambridge, MA 02138, USA}
\begin{abstract}

We provide a novel approach for characterization of quantum
Hamiltonian systems via utilizing a single measurement device.
Specifically, we demonstrate how external quantum correlations can
be used for Hamiltonian identification tasks. We explicitly
introduce experimental procedures for direct estimation of single-
and two-qubit Hamiltonian parameters, and also for simultaneous
estimation of transverse and longitudinal relaxation rates, using a
single Bell-state analyzer. An advantage of our method over the earlier approaches is that it has a built-in feature which makes it suitable for partial characterization of Hamiltonian parameters.


\end{abstract}
\maketitle

\section{Introduction}
Characterization of quantum dynamical systems is one of the most
fundamental problems in quantum physics, and lies at the heart of
quantum information processing \cite{nielsen-book} and coherent
control \cite{qcontrol}. There are a few known methods to achieve
this goal, such as standard quantum process tomography \cite{sqpt},
ancilla-assisted process tomography \cite{aapt}, and direct
characterization of quantum dynamics (DCQD)
\cite{dcqd,experiment1,experiment2}. Since the required physical resources
grow exponentially with the number of degrees of freedom, all
quantum process tomography schemes are in principle inefficient.
However, in various physical situations a full characterization of
the quantum dynamical superoperator is not always necessary, as
sometimes the information about relevant physical quantities could
be related to only a polynomial number of parameters in the system
size \cite{Emerson}. This is indeed the case when: (i) important
physical properties of a quantum system can be directly associated
only to a subset of certain superoperator elements, (ii) \textit{a
priori} knowledge exists about general properties of quantum
dynamics, and (iii) neglecting some elements will lead to small
system characterization errors.

The task of Hamiltonian identification, as a characterization of
quantum systems, is of paramount importance in quantum physics and
chemistry. In particular, it is required for monitoring or
controlling performance of noisy single- and two-qubit quantum
gates/devices in quantum information processing
\cite{nielsen-book,Mikko}. For various physical systems, a generic
form of the Hamiltonian can be guessed from general
physical/engineering considerations or observations. However, one
still needs to estimate the Hamiltonian parameters for a given
quantum system in order to study the internal dynamics of the system
and also to investigate the exact form and the strength of a
potential system-bath coupling.

Identification of time-independent (or piece-wise constant)
Hamiltonians along with the estimation of errors have already been
studied for both single-qubit and two-qubit cases
\cite{Cole-Hsinglequbit05,Devitt-Hg-twoqubit06}. Characterization of
a single-qubit Hamiltonian is achieved via determination of the
measurement results using a single fixed readout process which is a
periodic function of time. Through Fourier analysis of this signal
and other related techniques, identification is reduced to finding
the (relative) location of peaks and their heights of the Fourier
spectrum \cite{Cole-Hsinglequbit05}. Similarly, in the two-qubit
case, Hamiltonian parameters are obtained through some entanglement
measurements sampled in many time points, and then a Fourier
analysis determines the parameters \cite{Devitt-Hg-twoqubit06}.
There are also more general Hamiltonian identification schemes which
employ closed-loop learning control techniques along with efficient
and improved global laboratory data inversion for identification
\cite{GeremiaRabitz02}. These techniques are useful particularly
when one has access to tailored control fields (e.g., shaped laser
pulses) while the measurements are being performed. A fundamental
relevant question is how one can exploit external quantum
correlations in order to enhance identification of quantum
Hamiltonian systems. This is the subject we address in this work.

In this paper, we introduce an 
analytical method for direct
characterization of important classes of Hamiltonians. This method
is based on a newly proposed DCQD scheme \cite{dcqd}. In particular,
we demonstrate how to estimate all parameters of a general
time-independent single-qubit Hamiltonian and two-qubit (isotropic
or anisotropic) exchange Hamiltonian. A distinctive feature of our
method is that, when only some partial knowledge about the system is
requested, it does not require Fourier analysis of the experimental
data. In principle, this obviates the need for long sampling times
and in turn offers more controllability for the related estimation
process. Our Hamiltonian identification method is applicable to
quantum systems enabling two-body measurements, due to the fact that
DCQD requires Bell-state measurements (BSMs). The required BSM can
be in principle achieved in linear optical systems via
postselections \cite{White05} or hyperentanglement \cite{Kwiat}, and
also in trapped ions  \cite{iontrap} (see Ref.~\cite{iontrap2} for a deterministic,
programmed generation of ``ultralong lifetime" Bell-states),  and optical lattices
\cite{opticallattices}. In solid-state systems, several schemes for
controllable two-body interactions have been proposed
\cite{Emerson,Loss:98Kane:98Vrijen:00,fault,Hall,Imamoglu:99,exciton,Platzman:99},
with the state of the art experimental realization in semiconductor
quantum dots \cite{Petta05} and superconducting flux qubits \cite{niskanen}.

The evolution of a $d$-dimensional quantum system (open or closed)
with initial state $\rho$ can, under some natural assumptions, be
expressed in terms of a completely positive quantum dynamical map
$\mathcal{E}_{t}$, which can be represented as:
$\mathcal{E}_{t}(\rho)= \sum_{ij}\chi_{ij}(t)\sigma_{i}\rho
\sigma^{\dag}_j$. The positive matrix $\bm{\chi}=[\chi_{mn}]$
encompasses all information about the dynamics, relative to the
fixed operator basis set $\{\sigma_{m}\}$, where
$\text{tr}(\sigma_m^{\dag}\sigma_n)=d\delta_{mn}$. The theory of
DCQD determines elements of $\bm{\chi}$ matrix \cite{dcqd} by
relating them to measurement results more directly than the other
existing schemes. The main idea of DCQD is based on quantum error
detection theory in which by preparation of suitable states and
measurement of their stabilizers and normalizers partial information
about errors can be obtained. The required stabilizer and normalizer
measurement can be physically realized with a single Bell-state
analyzer. Table~\ref{dcqd-tab} summarizes the scheme for the
single-qubit case.


\begingroup
\squeezetable
\begin{table}[bp]
\begin{ruledtabular}
\caption{Input states and measurements for direct
characterization of single-qubit $\bm{\chi}$. Here
$|\Phi^{+}_{\alpha}\rangle=\alpha|00\rangle+\beta|11\rangle$,
$|\Phi^{+}_{\alpha}\rangle_{x(y)}=\alpha|++\rangle_{x(y)}+\beta|--\rangle_{x(y)}$
where $|\alpha |\neq |\beta |\neq 0$ and $\mathrm{Im}(\bar{\alpha}
\beta)\neq 0$, and $\{|0\rangle,|1\rangle \}$, $\{|\pm\rangle_x \}$,
$\{|\pm\rangle_y\}$ are eigenstates of the Pauli operators
$\sigma_z$, $\sigma_x$, and $\sigma_{y}$. $P_{\Phi^+}$ is the projector on the Bell state $|\Phi^+\rangle$, and similarly for the other projectors.}
\begin{tabular}{l|c|c}
input state & Bell-state measurement & output $\chi_{mn}$\\
 \colrule
$|\Phi^+\rangle$  &$P_{\Psi^{\pm}},P_{\Phi^{\pm}}$ &
$\chi_{00},\chi_{11},\chi_{22},\chi_{33}$\\
$|\Phi^+_{\alpha}\rangle$ & $P_{\Phi^{+}}
\pm P_{\Phi^{-}}, P_{\Psi^{+}}\pm P_{\Psi^{-}}$ & $\chi_{03},\chi_{12}$\\
$|\Phi^+_{\alpha}\rangle_x$ & $P_{\Phi^{+}}\pm P_{\Psi^{+}},
P_{\Phi^{-}}\pm P_{\Psi^{-}}$ & $\chi_{01},\chi_{23}$\\
$|\Phi^+_{\alpha}\rangle_y$  & $P_{\Phi^{+}}\pm P_{\Psi^{-}},
P_{\Phi^{-}}\pm P_{\Psi^{+}}$ & $\chi_{02},\chi_{13}$\\
\end{tabular}\label{dcqd-tab}
\end{ruledtabular}
\end{table}
\endgroup

\section{Identification of single-qubit Hamiltonians}
Let us consider the cases that quantum dynamics is generated by a
time-independent Hamiltonian,
$\mathcal{E}_{t}(\rho)=\mathcal{U}(t)^{\dag}\rho\ \mathcal{U}(t)$,
where $\mathcal{U}(t)=e^{-itH}$ ($\hbar\equiv1$), one obtains
$\chi_{mn}=a_m\bar{a}_n$, where $\mathcal{U}(t)=\sum_m
a_{m}(t)\sigma_m$ and $H$ is the Hamiltonian of the system. Since an
energy shift is always possible, we only consider traceless
Hamiltonians. In the single-qubit case, where
$H=\bm{J}\cdot\bm{\sigma}$, with the choice
of $\{\bm{\sigma}=(\sigma_x,\sigma_y,\sigma_z),\openone\}$ as the operator 
basis, we have: $\chi_{00}=c^{2}$, $\chi_{\alpha \alpha
}=s^{2}\hat{J}_{\alpha }^{2}$, $\chi_{0\alpha }=isc\hat{J}_{\alpha
}$, and $\chi_{\alpha \beta}=s^{2}\hat{J}_{\alpha }\hat{J}_{\beta
}$, in which $\bm{J}=J\hat{\bm{J}}$ ($J=\|\bm{J}\|$), $c=\cos (Jt)$,
$s=\sin (Jt)$ and $\alpha ,\beta =x,y,z$.

In order to find the real vector $\bm{J}$, according to the DCQD
theory, we can choose different experimental configurations
(measurement settings) depending on our \textit{a priori}
information about the Hamiltonian. If the signs of the Hamiltonian
parameters (i.e., the components of $\bm{J}$) are already known, we
can determine $J$ and the absolute values of the components,
$|\hat{J}_{\alpha}|$, in a single experimental configuration. First,
we prepare a maximally entangled state between the qubit of interest
$A$, and an ancilla, $B$, as $|\Phi^+ \rangle_{AB}
=(|00\rangle+|11\rangle )/\sqrt{2}$. Next, the system evolves under
$H$ for a duration of time $t$. Finally, we perform a BSM
represented by the four projection operators $P_{\Phi^{\pm }}$ and
$P_{\Psi^{\pm }}$. The probabilities of
obtaining these outcomes are found as
$\mathrm{tr}[P_{I}\mathcal{E}_{t}(\rho )]=c^2$ and
$\mathrm{tr}[P_{\alpha }\mathcal{E}_{t}(\rho )]=s^2\hat{J}_{\alpha
}^{2}$, where the projection operators $P_{I}$ and $P_{\alpha }$
(for $\alpha=x,y,z$) correspond to the states $\Phi ^{+}$,
$\Psi^{+}$, $\Psi^{-}$, and $\Phi^{-}$, respectively. Thus, we have
\begin{eqnarray}
&\hat{J}_{\alpha}^{~~2}=\mathrm{tr}[P_{\alpha }\mathcal{E}_{t}(\rho
)]/(1- \mathrm{tr}[P_{I}\mathcal{E}_{t}(\rho )]).
\label{H-tind-singlequbit}
\end{eqnarray}
The diagonal elements of the superoperator give the absolute values
of the unknown parameters $\hat{J}_{\alpha}$.
Equation~(\ref{H-tind-singlequbit}) bears this interesting result that
measurements at a single time-point $t$ are in principle (ignoring
the inherent issue of statistical errors) enough to obtain
$(|\hat{J}_x|,|\hat{J}_y|,|\hat{J}_z|)$. When the relative signs are
already known, this uniquely identifies the reference frame of the
Hamiltonian.

In order to obtain $J$, we are required to estimate the frequency of
the function
$\cos^2(Jt)=\text{tr}\left[P_I\mathcal{E}_t(\rho)\right]$. The
theory of signal processing and discrete Fourier analysis state that
one generally needs to perform many time samplings to obtain
frequencies. By the Nyquist criterion, the sampling frequency
$f_S\equiv 1/\tau_S$ must be bounded below by half of the frequency
of the original signal, i.e., $f_S>J$, to reduce the inherent
aliasing \cite{signal processing}. In
Refs.~\cite{Cole-Hsinglequbit05,Devitt-Hg-twoqubit06} one can find
more detailed analysis of these issues and how to read $J$ from
experimental data. Specifically, in
Ref.~\cite[(b)]{Devitt-Hg-twoqubit06} an interesting method of
ensemble measurements in sample points has been introduced that can
reduce the statistical error in inference.

In the more general case, to fully characterize the real vector
$\bm{J}$ we need to consider a different strategy and perform two
measurements for the off-diagonal elements of $\bm{\chi}$. According
to DCQD, these two experimental configurations are sufficient to
determine the diagonal of the superoperator, $\chi _{ii}$ for $i\in
\{0,1,2,3\}$, and four off-diagonal parameters $\text{Im}(\chi
_{0i})$, and $\text{Re}(\chi_{jk})$, for any two sets of values of
$\{i,j,k\},\{i',j',k'\}\in \{1,2,3\}$ such that $i\neq i'$, $j\neq
k\neq i$ and $j'\neq k'\neq i'$. For example, by preparation of a
nonmaximally entangled state $|\Phi^+ _{\alpha}\rangle =\alpha
|00\rangle +\beta |11\rangle $ (Table~\ref{dcqd-tab}) and performing a
standard BSM, we can obtain the following equations:
\begin{eqnarray*}
&\chi_{00}+\chi_{33}=p_{+}, ~~\chi _{11}+\chi _{22}=p_{-}, \\
&a(\chi_{00}-\chi_{33})+b\mathrm{Im}(\chi _{03})=c_{+}, \\
&a(\chi_{11}-\chi_{22})-b\mathrm{Re}(\chi _{12})=c_{-},
\end{eqnarray*}
with $p_{\pm }=\mathrm{tr}[P_{\pm 1}\mathcal{E}_{t}(\rho )]$,
$a=\mathrm{tr}(\mathcal{N}\rho )$, $b=2i\mathrm{tr}(\sigma
_{z}^{A}\mathcal{N}\rho )$, and $c_{\pm }=p_{\pm
}\mathrm{tr}(\mathcal{N}\rho _{\pm 1})$, where $\rho =|\Phi^+
_{\alpha}\rangle \langle \Phi^+ _{\alpha}|$,
$\mathcal{N}=\sigma_x^A\sigma_x^B$, $P_{+1}=P_{\Phi^+}+P_{\Phi^-}$,
$P_{-1}= P_{\Psi^+}+P_{\Psi^-}$ and $\rho _{\pm 1}=P_{\pm
1}\mathcal{E}_{t}(\rho )P_{\pm 1}/\mathrm{tr}[P_{\pm
1}\mathcal{E}_{t}(\rho )]$. In the other experimental configuration,
we prepare a nonmaximally entangled state
$|\Phi^+_{\alpha}\rangle_x$ and perform another standard BSM to
obtain a similar set of equations in the $\{|\pm \rangle _{x}\}$
basis (here and also for $\{|\pm\rangle_y\}$ basis,
$\mathcal{N}=\sigma_z^A\sigma_z^B$). Using these linearly
independent equations we can determine diagonal elements of the
superoperator, $\chi _{ii}$ ($i=0,1,2,3$) and four off-diagonal
parameters $\text{Im}(\chi_{03})$, $\text{Im}(\chi _{01})$,
$\text{Re}(\chi _{12})$, and $\text{Re}(\chi_{23})$. As we have
shown above, the diagonal elements can be used to determine $J$ and
the absolute values $|\hat{J}_{\alpha}|$. The relative signs of
$\hat{J}_{x}$, $\hat{J}_{y}$, and $\hat{J}_{z}$ can be found from
the off-diagonal parameters above; so, we can identify $\bm{J}$ up
to a global sign. This global sign is usually evident from the
physical/engineered setup under consideration, e.g., from the
direction of a global magnetic field for spin systems. In physical
situations where this global sign cannot be deduced from general
physical considerations, we need to perform a third measurement that
corresponds to characterizing $\text{Im}(\chi _{02})$ and
$\text{Re}(\chi_{31})$ which completes our knowledge about an
arbitrary (time-independent) single-qubit Hamiltonian. The whole
analysis is also applicable to the case of piece-wise constant
Hamiltonians. In the following we discuss another important example
of single-qubit dynamics, although non-Hamiltonian, that shows how
the DCQD estimation may provide advantage in estimation of dynamical
parameters in the Markovian regime.

\section{Simultaneous determination of $T_{1}$ and $T_{2}$}

Let us consider the so-called quantum \textit{homogenization}
process acting on a single-qubit density matrix $\rho(0)$ for time
$t$, where $\rho_{00}(0)=a$ and $\rho_{01}(0)=b$ in the
$\{|0\rangle,|1\rangle\}$ basis. This leads to the final state
$\rho(t)$ with
$\rho_{00}(t)=(a-a_{\infty})\exp(-t/T_{1})+a_{\infty}$ and
$\rho_{01}(t)=b\exp(-t/T_{2})$, where $a_{\infty}$ characterizes the
population of thermal equilibrium state, and the time-scales $
T_{1}$ and $T_{2}$ ($T_{2}\leqslant 2T_{1}$) are longitudinal and
transverse relaxation time-scales of the system, respectively
\cite{nielsen-book,scarani}. 
The explicit form of $\chi_{ii}$ elements are as follows: $\chi_{00(33)}=[\exp(-t/T_1)\pm 2\exp(t/T_2)+1]/4$,
$\chi_{11}=\chi_{22}=[\exp(t/T_1)+1]/4$.

Now we demonstrate that both $T_{1}$ and $T_{2}$ can always be
estimated in a \textit{single} ensemble measurement by using the
DCQD scheme for estimating diagonal elements of $\bm{\chi}$. We
first prepare a Bell-state $|\Phi^+\rangle_{AB}$, 
and then let the qubit $A$ interact with a thermalizing environment
for a given time $t$. The outcomes of a BSM yield the
following relations for $T_1$ and $T_2$:
\begin{eqnarray}
\aligned &1/T_{1}=-\ln\left(
2\mathrm{tr}[P_{\Psi^{+}}\mathcal{E}_{t}(\rho
)]+2\mathrm{tr}[P_{\Psi^{-}}\mathcal{E}_{t}(\rho
)] -1\right)/t,\\
&1/T_{2}=-\ln\left(\mathrm{tr}[P_{\Phi^+}\mathcal{E}_t(\rho)]-
\text{tr}[P_{\Phi^-}\mathcal{E}_t(\rho)]\right)/t.
\endaligned
\label{t1t2}
\end{eqnarray}
Ideally, these equations imply adequacy of single time-point
measurements. That is, unlike the case of reading $J$, where time
sampling is necessary and aliasing is inevitable, $T_1$ and $T_2$
can in principle be obtained through single time-point measurements.
This feature could be utilized to reveal the non-Markovian nature of
system-bath interaction; e.g., by observing time variations in the
estimated relaxation and dephasing rates beyond the the natural
deviation due to statistical errors. Moreover, due to orthogonality
of BSM outcomes, it is easy to unambiguously distinguish $T_1$ from
$T_2$, unlike the approach presented in Ref.~\cite{ColePRA06}.
Traditionally, in order to measure the longitudinal and transverse
relaxation times, one needs to measure two non-commutative
observables (e.g., Pauli operators $\sigma_z$ and $\sigma_x$) on two
subensembles of identical systems, corresponding to magnetization
vectors $M_{z}$ and $M_{xy}$ parallel and perpendicular to a global
magnetic field $B_{0}$. The number of repetitions in each
measurement is determined by the desired accuracy in the time
sampling estimation of the relaxation times associated with
magnetizations $M_{z}$ and $M_{xy}$ \cite{NMR}.

\section{Identification of Two-qubit exchange Hamiltonians}

In solid-state systems, it is often the case that each pair of
qubits ($AB$) interact directly or effectively through an exchange
Hamiltonian $H_{\text{ex}}=\sum_{\alpha}J_{\alpha}\sigma
_{\alpha}^{A}\sigma_{\alpha}^{B}$, where $J_{\alpha}$s are the
couplings of the two-qubit interaction (see also
Ref.~\cite{opticallattices} for the exchange interaction between
neutral atoms in optical lattices). The case of isotropic or
Heisenberg interaction ($J_{x}=J_{y}=J_{z}$) is intrinsic to
spin-coupled quantum dots, and donor atom nuclear/electron spins
\cite{Loss:98Kane:98Vrijen:00}. This interaction is also important
in the context of universal fault-tolerant quantum computing
\cite{fault}.
The XY Hamiltonian ($J_x=J_y$, $J_{z}=0$) is the available
interaction in quantum Hall systems \cite{Hall}, quantum dots/atoms
in cavities \cite{Imamoglu:99}, and exciton-coupled quantum dots
\cite{exciton}. The XXZ ($J_x=J_y\neq J_{z}\neq 0$) interaction
appears in the electrons in liquid-Helium quantum computing
proposals \cite{Platzman:99}.

In the case of XYZ Hamiltonians, 
the $\bm{\chi}$ matrix has only 10 nonzero independent elements
$\chi_{mn}$, for $m,n=0,5,10,15$. Similar to the case of the general
single-qubit Hamiltonian, these diagonal elements contain information only
about the absolute values $|J_{\alpha}|$s.
In order to obtain information about the signs of $J_{\alpha}$s, we
need to measure off-diagonal elements as well. However, in most
physical/practical cases the signs of the terms in an exchange
Hamiltonian are already known from some general properties of the
system. For example, for many materials it is known whether below
the phase transition point they become ferromagnetic or
anti-ferromagnetic---alternatively this information can be obtained
for a given material by measuring its linear response to an
applied magnetic field. In these cases, the Hamiltonian can be
completely determined with a \textit{single} ensemble measurement
corresponding to the diagonal elements of the superoperator.

Let us consider the important classes of isotropic and anisotropic
exchange interactions. For these Hamiltonians the sign of $J$ is
known from the ferromagnetic property of the system. In fact, by
definition $J=E_{S}-E_{T}$ (where $E_{S}$ and $E_{T}$ are the energy
of singlet and triplet states), is always negative for ferromagnetic
materials. For example, for a two-electron system, the singlet state
is the ground state of the system if $J<0$. On the contrary, for
anti-ferromagnetic materials, $J$ is always positive which indicates
that in the ground state spins tend to arrange themselves in the
same direction.


In order to determine the diagonal elements $\chi_{ii}$ for
Heisenberg interaction between two electrons $A_{1}$ and $A_{2}$,
one can prepare a tensor product of maximally entangled states
between each electron and a pair of ancilla electrons ($B_{1}$ and
$B_{2}$) such as
$|\Phi^{+}\rangle_{A_1B_1}|\Phi^{+}\rangle_{A_2B_2}$. Then, the
unknown Hamiltonian $H$ for the duration of $t$ is applied, and a
tensor product of BSMs acting on each pair $A_i B_i$ is performed,
where this operation can be represented by a tensor product of
$P_{\Phi_{i}^{+}}$, $P_{\Psi_{i}^{+}}$, $P_{\Psi_{i}^{-}}$,
$P_{\Phi_{i}^{-}}$ for $i=1,2$. The joint probability distributions
of the BSMs are related to $J$ through $\mathrm{tr}[P_{\Phi
_{1}^{+}}P_{\Phi _{2}^{+}}\mathcal{E}_{t}(\rho )]=c^6+s^6$ and
$\mathrm{tr}[P_{\Psi _{1}^{+}}P_{\Psi _{2}^{+}}\mathcal{E}_{t}(\rho
)]=\mathrm{tr}[P_{\Psi _{1}^{-}}P_{\Psi
_{2}^{-}}\mathcal{E}_{t}(\rho )]=\mathrm{tr}[P_{\Phi
_{1}^{-}}P_{\Phi _{2}^{-}}\mathcal{E}_{t}(\rho )]=s^2c^2$.
Therefore, we have:
\begin{eqnarray}
&\sin(2|J|t)=2\sqrt{\mathrm{tr}[P_{\Phi _{1}^{-}}P_{\Phi
_{2}^{-}}\mathcal{E}_{t}(\rho )]}, \label{Eq-isoH}
\end{eqnarray}
and similar relations hold for $P_{\Psi _{1}^{+}}P_{\Psi _{2}^{+}}$
and $P_{\Psi _{1}^{-}}P_{\Psi _{2}^{-}}$ as well.

In the case of anisotropic exchange interactions one can perform a
similar Bell-state preparation and BSM as in the case of isotropic
exchange, to obtain:
\begin{eqnarray}
&\sin(2|J_{x}|t)=2\sqrt{\mathrm{tr}[P_{\Psi _{1}^{\pm }}P_{\Psi
_{2}^{\pm }}\mathcal{E}_{t}(\rho )]},  \label{Eq-anisoH1}\\
&\hskip -3.5mm \cos(2|J_{z}|t)=\sqrt{(\mathrm{tr}[P_{\Phi
_{1}^{+}}P_{\Phi _{2}^{+}}\mathcal{E}_{t}(\rho
)]-s_x^4)/(c_x^4-s_x^4)}. \label{Eq-anisoH2}
\end{eqnarray}
To read $|J_{\alpha}|$, one needs to have time samplings (i.e.,
ensemble measurements for many time-points) and follow the Fourier
analysis based method sketched earlier. Therefore, having \textit{a
priori} knowledge about the ferromagnetic property of the system,
one can identify the underlying exchange Hamiltonian.

Note that the energy spectrum of $H_{\text{ex}}$ can be simply
calculated using the above relations and knowing the fact that
Bell-states are the eigenkets of the exchange Hamiltonian.
Eigenvalues of $H_{\text{ex}}$ can be written as $E=\pm |J_{\alpha
}|\pm |J_{\beta }-J_{\gamma }|$, where $\alpha \neq \beta \neq
\gamma \in \{x,y,z\}$. We have already shown how to estimate
$|J_{\alpha}|$ for $\alpha=x,y,z$. In order to find relative signs
of any two other components, such as $|J_{y}-J_{z}|$, the DCQD
algorithm can be utilized by performing a single ensemble
measurement that corresponds to measuring the off-diagonal element
$\chi _{0,5}$. For full characterization of an exchange Hamiltonian
without having any \textit{a priori} knowledge about the signs of
the coupling constants, one needs to measure the off-diagonal
element $\chi _{0,10}$ too. Therefore, with a total of three
ensemble measurements, corresponding to $\chi _{i,i}$, $\chi_{0,5}$,
and $\chi_{0,10}$, full characterization of $H_{\text{ex}}$ can be
achieved.

\section{Remarks on precision}

In a realistic estimation process, due to decoherence, limited
measurement or preparation accuracies (because of specific
device architecture or finite ensemble
size), and other imperfections, some
errors may occur (a generalization of the DCQD theory that addresses faluty preparations 
and measurements is underway and will
be reported elsewhere \cite{mohseni-rezakhani}). These factors might affect the amount of physical resources required for 
a given accuracy of the estimation, hence some appealing features like sufficiency of single-time 
point measurements might be lost. For the case of ideal
preparations/measurmens scenario in which single time-point
measurements are in principle sufficient, errors scale up as $1/\sqrt{N_E}$,
where $N_E$ is the number of repeated measurements. In the cases we need to perform
time samplings, the error in the estimation of frequencies (and
thus, Hamiltonian parameters) is governed by the Nyquist criterion
and the quantum shot-noise limit \cite{huelga}. Let us consider
$N_S$ samples, for each of which we perform $N_E$ measurements.
Therefore, according to the quantum shot-noise limit, $\Delta f\sim
1/(N_S\tau_S\sqrt{N_E})$, and the Nyquist criterion, $ f_S=1/\tau_S
\geqslant f/2$, we get: $\Delta f/f\sim 1/(N_S\sqrt{N_E})$ (see
Ref.~\cite[(b)]{Devitt-Hg-twoqubit06}). That is, the average error
in estimation of Hamiltonian parameters scales as
$1/(N_S\sqrt{N_E})$. In other words, for an error $\epsilon$, or
with the number of digits of precision $\log(1/\epsilon)$, we need
$\text{poly}(1/\epsilon)$ more steps, which is common among all
Fourier analysis based methods \cite{brown}. In this respect, our
method does not provide an advantage over the one in
Refs.~\cite{Devitt-Hg-twoqubit06}---both methods provide similar
accuracy scaleup. However, the advantage of our method lies in the
built-in ability of the method for partial characterization. That
is, there is a level of independency in the way different sets of
parameters are related to measurements results. For more general
discussions on partial characterization by DCQD see
Refs.~\cite{dcqd} and for a very recent proof-of-principle
experiment on this issue, using polarization and spatial degrees of
freedom of a single photon, see Ref.~\cite{experiment2}. Moreover,
in our method some of the parameters are related more directly to
the measurements data, hence obviating the need to (a full)
inversion in the first place. E.g., in the single-qubit case, we
obtain $|\hat{J}_{x,y,z}|$ just by a very simple algebraic
manipulation of the data. This feature is not necessarily available
in the methods of Refs.~\cite{Devitt-Hg-twoqubit06}, because a
Fourier analysis would be necessary even for extracting a partial
information about the Hamiltonian.

\section{Conclusion and discussion}

We have presented a theoretical approach for utilizing auxiliary
quantum correlations to perform Hamiltonian identification. In this
method one can directly obtain full information about unknown
parameters of time-independent of single- and two-qubit Hamiltonians
without full quantum process tomography. In addition, we demonstrate
that for a single qubit undergoing a generic Markovian homogenizing
quantum map, both related relaxation times can be estimated
simultaneously by utilizing a single Bell-state measurement.
Furthermore, we illustrate how our prior knowledge about Hamiltonian
systems can be exploited in order to reduce the required physical
resources for identification tasks. In particular, we show that the
required repeated measurements, associated to time sampling of data,
can be reduced when we are interested in partial characterization of
the Hamiltonian systems and also for estimating relaxation rates. 
With the recent advent of various
methods for generation of controllable entanglement, our proposed
method may have near-term application to a variety of quantum
systems/devices, such as in trapped ions, liquid-state NMR, optical
lattices, and entangled pairs of photons.


Discussions with J. D. Biamonte, J. H. Cole, and D. A. Lidar are
acknowledged. This work was supported by NSERC (to M.M.), the
Faculty of Arts and Sciences of Harvard University (to M.M. and
A.A.), \textit{i}CORE, MITACS, and PIMS (to A.T.R.).


\end{document}